\documentclass[proceedings]{JHEP3} 
\PrHEP{hep2001}
\conference{International Europhysics Conference on High Energy Physics} 
\title{Extended Minimal Flavour Violating MSSM}
\author{\speaker{E.~Lunghi}\thanks{E.L. acknowledges financial support 
from the Alexander Von Humboldt Foundation.} \\ Deutsches
Elektronen Synchrotron, DESY, Notkestr. 85, D-22607 Hamburg,
Germany \\
E-mail: \email{lunghi@mail.desy.de}}

\usepackage{epsfig}

\def \ra{\rightarrow}

\def \beq{\begin{equation}}
\def \eeq{\end{equation}}
\def \bea{\begin{eqnarray}}
\def \eea{\end{eqnarray}}
\def \ben{\begin{enumerate}}
\def \een{\end{enumerate}}
\def \bit{\begin{itemize}}
\def \eit{\end{itemize}}

\def \branch{{\cal B}}
\def \gev{{\hbox{GeV}}}

\def \tev{{\hbox{TeV}}}

\def \cl#1{{#1\%\ \mathrm{C.L.}}}

\def \fig#1{Fig.~\ref{#1}}

\def \nn{\nonumber}

\def \a{\alpha}
\def \b{\beta}
\def \D{\Delta}
\def \g{\gamma}

\def \d{\delta}
\def \e{\epsilon}

\def \m{\mu}

\def \p{\pi}

\abstract{ The recently reported measurements of the CP asymmetry
$a_{\psi K}$ by the BABAR and BELLE collaborations are in good
agreement with the standard model prediction. With the anticipated
precision in $a_{\psi K}$ at the B factories and hadron colliders, one
hopes to pin down any possible deviation from the SM. We discuss an
extension of the MFV-supersymmetric models which comfortably
accommodates the current measurements of the CP asymmetry $a_{\psi
K}$, but differs from the SM due to an additional flavour changing
structure beyond the CKM matrix. We analyze the compatibility of this
model with present data and suggest specific tests in forthcoming
experiments in $B$ physics. In addition to the CP-asymmetries in
$B$-meson decays and the $B_s^0$ - $\overline{B_s^0}$ mass difference,
we emphasize measurements of the radiative transition $b \to d \gamma$
as sensitive probes of the postulated flavour changing structure.
Interestingly, the CKM--unitarity analysis in these models also allows
solutions with $\gamma > \pi/2$, where $\gamma=-\arg V_{ub}$. Such
large values of $\gamma$ are hinted by the current measurements of the
branchig ratios for the decays $B\to \pi\pi$ and $B\to K \pi$.  }

\begin{document}
\section{Outline of the model}
\label{outline}
This talk is based on the paper in Ref.~\cite{emfv}. The
supersymmetric model that we consider is essentially based on the
assumptions of heavy squarks (of the first two generations) and
gluinos. The charged Higgs and the lightest chargino and stop masses
are required to be heavier than $100 \; \gev$ in order to satisfy the
lower bounds from direct searches. The rest of the SUSY spectrum is
assumed to be almost degenerate and heavier than $1 \; \tev$.  In this
framework the lightest stop is almost right--handed and the stop
mixing angle (which parameterizes the amount of the left-handed stop
$\tilde t_L$ present in the lighter mass eigenstate) turns out to be
of order $O(M_W / M_{\tilde q}) \simeq 10\%$; for definiteness we will
take $|\theta_{\tilde t}| \leq \p /10$.

The assumption of a heavy gluino totally suppresses any possible
gluino--mediated SUSY contribution to low energy observables. On the
other hand, the presence of only a single light squark mass eigenstate
has strong consequences. In the MIA-framework~\cite{mia}, all
the FC effects which are not generated by the CKM mixing matrix are
proportional to the properly normalized off--diagonal elements of the
squark mass matrices.  In order to take into account the effect of a
light stop, we exactly diagonalize the $2\times 2$ stop system and
adopt the slightly different MIA implementation proposed in
Ref.~\cite{mia2}.  In this approach, a diagram can contribute sizably
only if the inserted mass insertions involve the light stop. All other
diagrams require necessarily a loop with at least two heavy squarks
and are therefore suppressed. This leaves us with only two
unsuppressed flavour changing sources other than the CKM matrix,
namely the mixings $\tilde u_L - \tilde t_2$ (denoted by $\d_{\tilde
u_L \tilde t_2}$) and $\tilde c_L - \tilde t_2$ (denoted by
$\d_{\tilde c_L \tilde t_2}$). 

In order to deal with a predictive model, we choose to set $\d_{\tilde
u_L \tilde t_2}=0$. This assumption is partly justified by the
remarkable agreement of the measured rate of the inclusive decay $B\to
X_s \g$ ($\branch^{b\to s\g} = (3.22 \pm 0.40) \times
10^{-4}$~\cite{cleo,aleph,belle}) with the SM prediction
($\branch^{b\to s\g}_{SM} = (3.29 \pm 0.33) \times
10^{-4}$~\cite{bsgnlo}). This hypothesis will be further tested in
CP-asymmetries ${\cal A}_{\rm CP}^{B \to (X_s,K^*) \gamma}$ at the
B-factories. Notice that the exclusion of $\d_{\tilde c_L \tilde t_2}$
from our analysis introduces strong correlations between the physics
that governs $b\ra d$ and $b\ra s$ transitions.

The free parameters of the model are the common mass of the heavy
squarks and gluino ($M_{\tilde q}$), the mass of the lightest stop
($M_{\tilde t_2}$), the stop mixing angle ($\theta_{\tilde t}$), the
ratio of the two Higgs vevs ($\tan \b_S$), the two parameters of the
chargino mass matrix ($\m$ and $M_2$), the charged Higgs mass
($M_{H^\pm}$) and $\d_{\tilde u_L \tilde t_2}$.  All these parameters
are assumed to be real with the only exception of the mass insertion
whose phase in not restricted {\it a priori}.

\section{SUSY contributions}
\label{susy}
The effective Hamiltonian that describes the $B_d - \bar B_d$ system
is
\begin{eqnarray}
    {\cal{H}}_{eff}^{\Delta B=2}
&=&
    -\frac{G_{F}^{2} M_{W}^{2}}{(2 \pi)^{2}} (V_{tb} V_{ts}^*)^{2}\ 
    \left( C_{1}\  \bar{d}_{L}\gamma^{\mu}
    b_{L}\cdot \bar{d}_{L}\gamma_{\mu}b_{L}\ 
    +\  C_{2}\  \bar{d}_{L}b_{R}\cdot 
    \bar{d}_{L} b_{R}
       \right. \nn \\
& &
       \left. 
    +\ C_3\ \bar{d}^{\alpha}_{L}b^{\beta}_{R}\cdot 
    \bar{d}^{\beta}_{L}b^{\alpha}_{R}\right) + {\it h. c.}\; ,
\end{eqnarray}
where $\alpha, \beta$ are colour indices. The $B_s$ and $K$
system cases are obtained respectively with the substitutions
$d\to s$ and $b\to s$.   

It is possible to show~\cite{emfv} that, in the framework described in
sec.~\ref{outline}, only $C_1$ receives sizable contributions. In
models in which the split between the two stop mass eigenstates is not
so marked, diagrams mediated by the exchange of both stops must be
considered and it is possible to find regions of the parameter space
(for large $\tan \b_S$) in which SUSY contributions to $C_3$ are
indeed dominant~\cite{oscar}. We note that the $\tan^4 \b_S$ enhanced
contributions to the coefficients $C_{2,3}$ whose presence is pointed
out in Ref.~\cite{janusz}, do not impact significantly for the range
of SUSY parameters that we consider ($\tan \b_S< 35$ and $|\theta_{
\tilde t}| < \pi/10$).

The structure of the SUSY contributions can be summarized as follows:
\bea 
\D M_{B_s}: & C_1^{SM} \rightarrow & C_1^{SM} (1+f) \nn \\ 
\e_K, \; \D M_{B_d}, \; a_{\psi K_S}: & C_1^{SM} \rightarrow &
             C_1^{SM} \left(1+f+ g \right) \nn 
\eea 
where 
\bea
f &\equiv& {(C_1^{H^\pm} + C_1^\chi) / C_1^{SM}} \; , \\
g &\equiv& g_R + i g_I \equiv {\overline C_1^{MI} \d_{\tilde u_L \tilde t_2}^2 / C_1^{SM}} \; .
\eea
$C_1^{H^\pm}$, $C_1^\chi$ and $\overline C_1^{MI} \d_{\tilde u_L
\tilde t_2}^2$ are the contributions of the charged Higgs and of
charginos without and with the mass insertion~\cite{emfv}. Note that
the only complex phase enters through $\d_{\tilde u_L \tilde t_2}$.

In order to understand the possible size of the above depicted SUSY
contributions we varied the input parameters over a reasonable range (
$\mu,M_2,M_{H^\pm} \in [100,1000] \gev$, $M_{\tilde t_2}\in
[100,600]\gev$, $\theta_{\tilde t} \in [-0.3,0.3]$, $\tan \b \in
[3,35]$) and included the constraints from $B\to X_s \g$ ($2.41 \leq
\branch^{b\to s\g}\times 10^4 \leq 4.02$ at $\cl{95}$), the anomalous magnetic
moment of the muon ($10 \leq \d a_{\mu^+}\times 10^{10} \leq 74 $ at
$\cl{95}$~\cite{gm2exp}) and $B\to \rho \g$ ($\branch^{B\to \rho \g}/
\branch^{B\to K^* \g} < 0.28$ at $\cl{90}$~\cite{bellebdg}). The
result of this analysis is presented in~\fig{fg}. Clearly $f$ and
$|g|$ are restricted to the range $f<0.4$, $|g|<2.0$. 
\EPSFIGURE[ht]{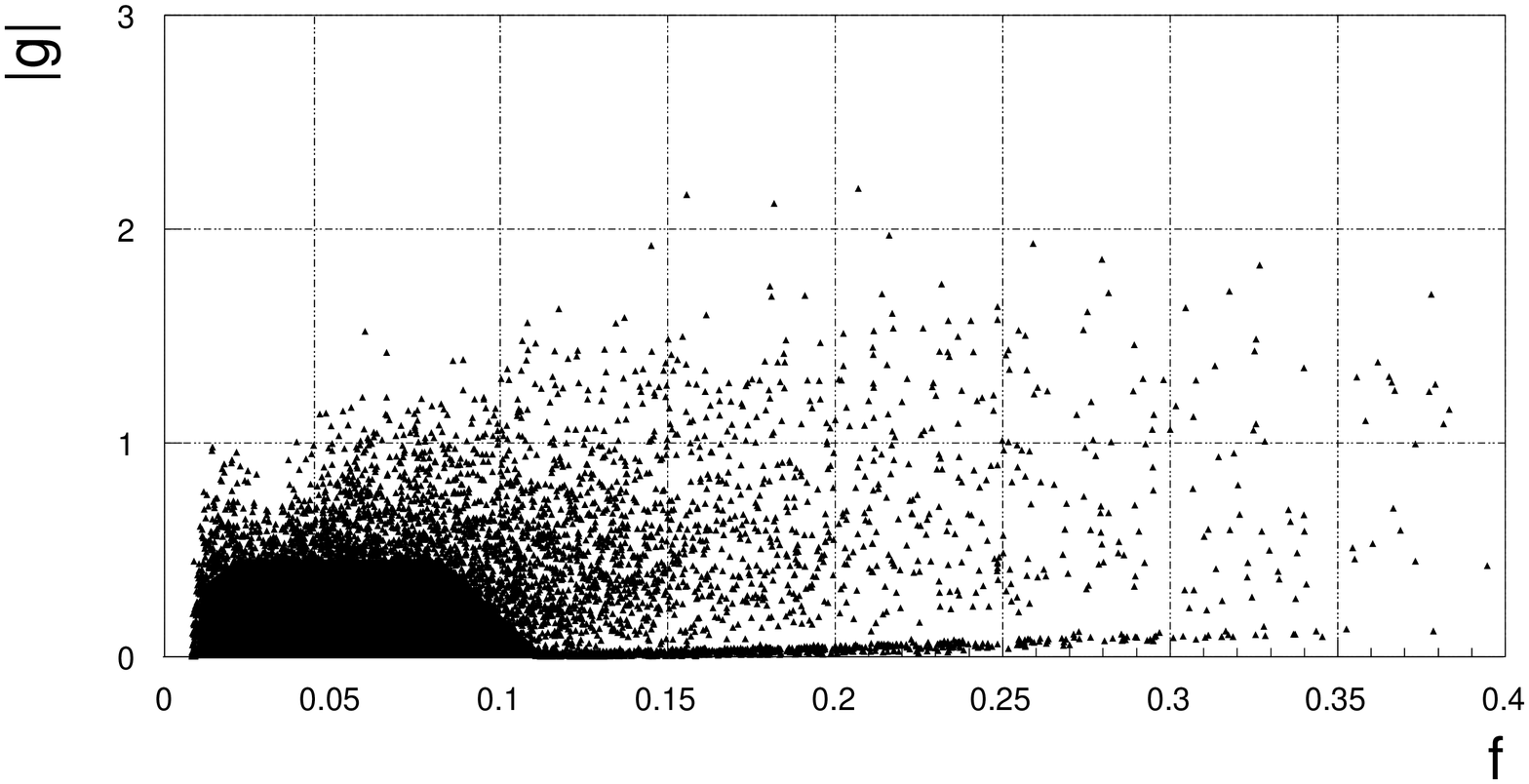,width=0.6\linewidth}{Allowed points in the 
$(f,|g|)$ plane. \label{fg}}

\section{Unitarity Triangle Analysis}
\FIGURE{
\epsfig{file=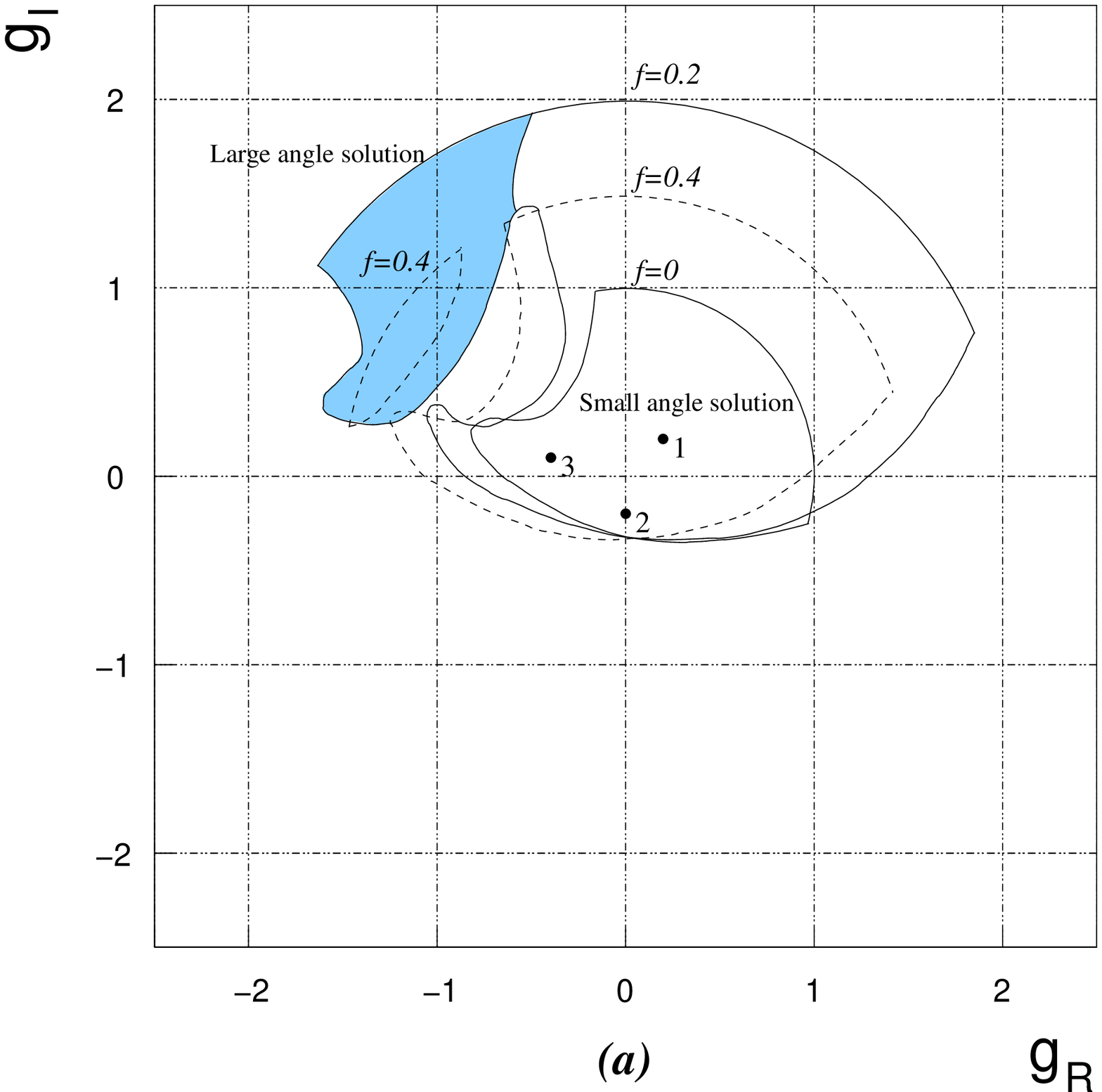,width=0.47\linewidth}
\epsfig{file=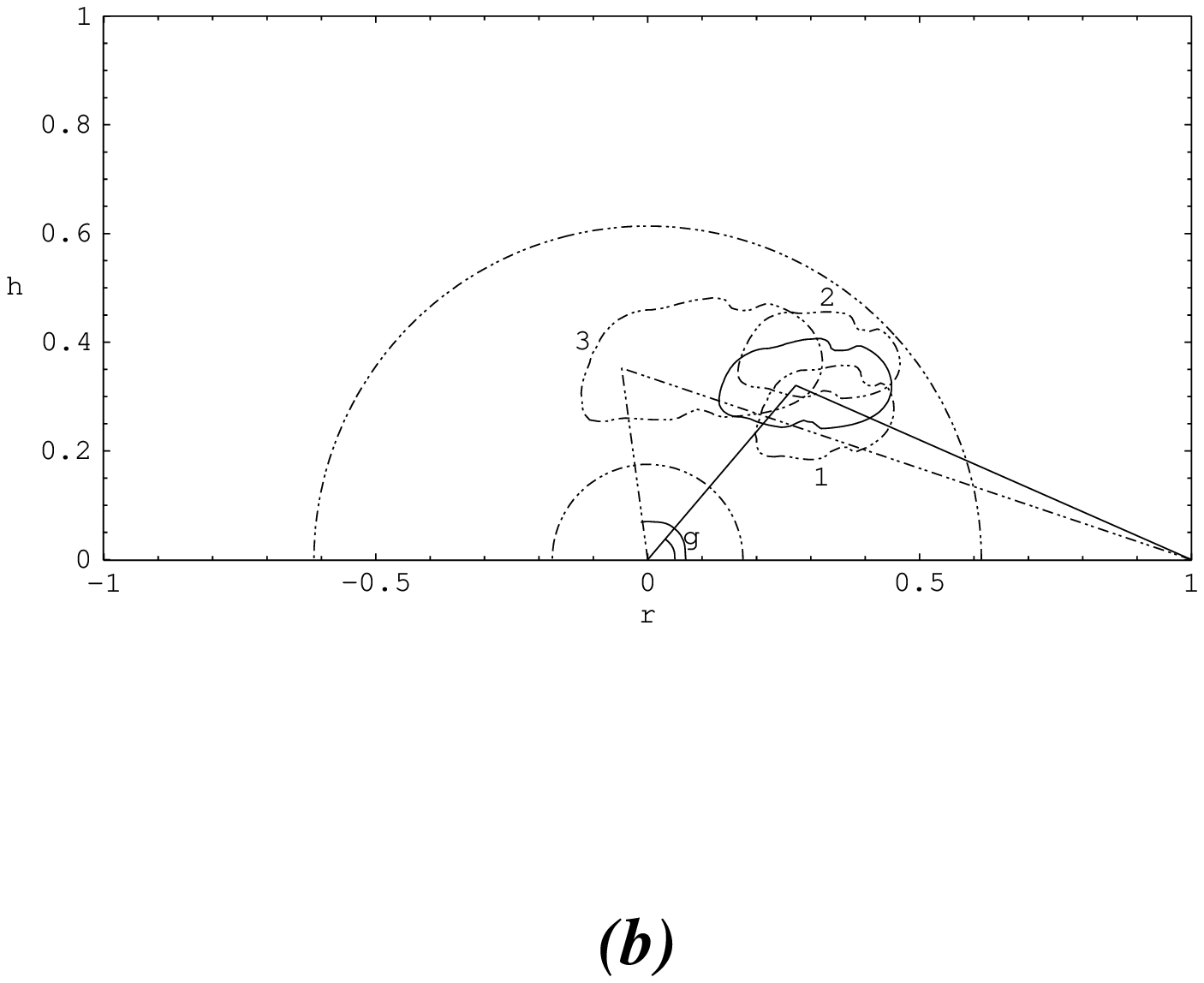,width=0.47\linewidth}
\caption{{\bf (a)} Region of the $(g_R , g_I)$ plane for which
$\min_{\rho,\eta} (\chi^2) \leq 2$ from the CKM-UT fits. 
The contours correspond to f = 0, 0.2 and 0.4 and the constraints on
$|g|$, coming from Fig.~\protect\ref{fg}, are $|g|\leq 1$, 2 and 1.5
respectively. {\bf (b)} Allowed $\cl{95}$ contours in the $(\bar\rho,\bar\eta)$
plane. The solid contour is the SM case. The two semicircles represent
the $2\; \sigma$ region allowed by $|V_{ub}/V_{cb}|=0.090\pm 0.025$.
The contours numbered 1 to 3 correspond respectively to the points
indicated in (a).}
\label{ampl}} 
In this section we analyze the implications of this parametrization on
the standard analysis of the UT (see Ref.~\cite{emfv} for a more
complete and extensive presentation).

Our first step is to investigate the regions of the parameter space
spanned by $f$, $g_R$ and $g_I$ that are favoured by the present
experimental data. The procedure consists in writing the $\chi^2$ of
the selected observables and in accepting only values of $f$ and $g$
which satisfy the condition $\min_{\rho,\eta} (\chi^2) \leq 2$.  The
resulting allowed regions are presented in \fig{ampl}a. The shaded
regions correspond to solutions in which the phase $\theta_d$ ( which
enters the analysis through $a_{\psi K_S}=\sin 2(\b+\theta_d)$ and is
given by $\theta_d=1/2 \arcsin (1+f+g)$) is extremely large. In this
case, large values of $|g|$ are required and these regions will be,
possibly, excluded once the lower bounds on sparticle masses and the
experimental errors on the branching ratio of $B\to X_s \g$ will
become more stringent. In the computation of $\chi^2$ we use
$\epsilon_K = (2.271 \pm 0.017) \; 10^{-3}$, $\D M_{B_d} = 0.484 \pm
0.010 \; {\rm ps}^{-1}$, $|V_{ub}/ V_{cb}| = 0.090 \pm 0.025$,
$a_{\psi K_s} = 0.79 \pm 0.12$~\cite{Aubert,Abashian} and $\D M_{B_s}
\geq 14.9 \; {\rm ps}^{-1}$.

\TABLE[ht]{
\begin{tabular}{|c|cc|c|c|c|cc|}
\hline 
Contour & $g_R$ & $g_I$ & $|V_{ub}/V_{cb}|$ & $\D M_{B_s}$ & $a_{\psi K_S}$ & $\a$ & $\g$ \cr \hline
1 & $0.2$  & $0.2$  & 0.094 & 20 $\hbox{ps}^{-1}$ & 0.78 & $119^\circ$ & $40^\circ$ \cr 
2 & $0.0$  & $-0.2$ & 0.110 & 20 $\hbox{ps}^{-1}$ & 0.71 & $101^\circ$ & $51^\circ$ \cr
3 & $-0.4$ & $0.1$  & 0.081 & 17 $\hbox{ps}^{-1}$ & 0.73 & $64^\circ$  & $98^\circ$ \cr \hline
\end{tabular}
\caption{\it Central values of the CKM ratio $|V_{ub}/V_{cb}|$, the
$B_s - \bar B_s$ mass difference, the $CP$ asymmetry $a_{\psi K_S}$
and the inner angles $\a$ and $\g$ of the unitarity triangle for the
contours of Fig.~\protect\ref{ampl}b.}
\label{values}}

In order to illustrate the possible different impact of this
parametrization on the unitarity triangle analysis, we focus on the
$f=0$ case and choose three generic points inside the allowed region
in the plane $(g_R,g_I)$.  In \fig{ampl}b we plot the $\cl{95}$
contours in the $(\bar \rho,\bar\eta)$ plane that correspond to the
points we explicitely show in \fig{ampl}a; the central values of the
various observables are summarized in Table~\ref{values}. Contour 3 is
particularly interesting since it corresponds to a solution in which
$a_{\psi K_S}$ is larger than in the SM and the Wolfenstein parameter
$\bar\rho$ is negative implying a value of the inner angle $\gamma$ in
the domain $\pi/2 < \gamma < \pi$. This is in contrast with the
SM--based analyses which currently lead to $\gamma<\pi/2$ at 2
standard deviations. We note that analyses~\cite{hou,beneke} of the
measured two--body non--leptonic decays $B\to \pi\pi$ and $B\to K \pi$
have a tendency to yield a value of $\gamma$ which lies in the range
$\gamma > \pi/2$ . While present data, and more importantly the
non--perturbative uncertainties in the underlying theoretical
framework do not allow to draw quantitative conclusions at present,
this may change in future. In case experimental and theoretical
progress in exclusive decays force a value of $\gamma$ in the domain
$\pi/2 < \gamma < \pi$, the extended--MFV model discussed here would
be greatly constrained and assume the role of a viable candidate to
the SM.

\section{Conclusions}
We investigated an extension of the so--called Minimal Flavour
Violating MSSM and its implications on $B$ physics. The non--CKM
structure in this Extended-MFV model reflects the two non--diagonal
mass insertions which influence the transitions $b \to d$ and $b \to
s$.  In the present analysis, we have neglected the effect of the mass
insertion in the $b \to s$ transition. We have shown that, as far as
the analysis of the unitarity triangle is concerned, it is possible to
encode, in this model, all SUSY effects in terms of three parameters
($f$, $g_R$ and $g_I$). We have worked out the allowed regions in the
plane $(f,|g|)$ by means of a scatter plot scanning the underlying
SUSY parameter space and taking into account constraints from $B\to
(X_s,\rho) \g$ and $(g-2)_\m$. We have then used these informations
and the requirement of compatibility with the fit of the unitarity
triangle in order to single out the overall allowed ranges of $f$,
$g_R$ and $g_I$. We have then chosen some generic points in order to
show the impact of these models on the phenomenology of the unitarity
triangle. Remarkably, we found solutions that admit $\g > \g^{SM}$ as
it is suggested by analyses of the decays $B\to K\pi$ and $B\to
\pi\pi$. It is also possible to show that the same SUSY parameter
space leaves room for sizable contributions to observable related to
the transition $B\to\rho \g$~\cite{emfv}. In particular, we considered
the ratio $\branch^{B\to \rho\g}/\branch^{B\to K^* \g}$, the isospin
breaking ratio (present since $\branch^{B^\pm\to \rho\pm \g}/2
\branch^{B^0\to \rho^0 \g}\neq 1$) and the $CP$ asymmetry
in the charged modes.


\begin{thebibliography}{100}

\bibitem{emfv}
A.~Ali and E.~Lunghi, hep-ph/0105200.

\bibitem{mia}
L.~J. Hall, V.~A. Kostelecky, and S.~Raby,
\newblock Nucl. Phys. {\bf B267}, 415 (1986).

\bibitem{mia2}
A.~J. Buras, A.~Romanino, and L.~Silvestrini,
\newblock Nucl. Phys. {\bf B520}, 3 (1998), [hep-ph/9712398].

\bibitem{cleo} 
D.~Cassel [CLEO Collaboration], talk presented at the
XX International Symposium on Lepton and Photon Interactions at High
Energies, Rome, Italy, Jul. 23-28, 2001. 

\bibitem{aleph}
R.~Barate {\em et~al.} [ALEPH Collaboration],
\newblock Phys. Lett. {\bf B429}, 169 (1998).

\bibitem{belle}
K.~Abe {\it et al.} [BELLE Collaboration],
Phys.\ Lett.\ B {\bf 511} (2001) 151
[hep-ex/0103042].

\bibitem{bsgnlo}
K.~Chetyrkin, M.~Misiak, and M.~M{\"u}nz,
\newblock Phys. Lett. {\bf B400}, 206 (1997), 
Erratum-ibid.\ B {\bf 425} (1997) 414, [hep-ph/9612313].

\bibitem{oscar}
D.~A. Demir, A.~Masiero, and O.~Vives,
\newblock Phys. Rev. Lett. {\bf 82}, 2447 (1999), [hep-ph/9812337].

\bibitem{janusz}
A.~J.~Buras, P.~H.~Chankowski, J.~Rosiek and L.~Slawianowska, hep-ph/0107048.

\bibitem{gm2exp}
H.~N. Brown {\em et~al.} [Muon g-2 Collaboration],
\newblock Phys. Rev. Lett. {\bf 86}, 2227 (2001), [hep-ex/0102017].

\bibitem{bellebdg}
A.~Abashian {\em et~al.} [BELLE Collaboration],
\newblock BELLE-CONF-0003  (2000).

\bibitem{Aubert}
B.~Aubert {\em et~al.} [BABAR Collaboration],
\newblock hep-ex/0107013.

\bibitem{Abashian}
K.~Abe {\em et~al.} [BELLE Collaboration],
\newblock hep-ex/0107061.

\bibitem{hou}
W.-S. Hou and K.-C. Yang,
\newblock Phys. Rev. Lett. {\bf 84}, 4806 (2000), [hep-ph/9911528].

\bibitem{beneke}
M.~Beneke, G.~Buchalla, M.~Neubert, and C.~T. Sachrajda,
\newblock (2001), hep-ph/0104110.

\end{thebibliography}
\end{document}